\documentstyle[prl,aps,preprint]{revtex}
\begin{document}
\tighten

\title{PURELY-NONPERTURBATIVE COMPOSITE OPERATORS  \\
AND PARTON DISTRIBUTIONS}
\author{{Xiangdong Ji}
\thanks{On leave of absence from Department of Physics, MIT, Cambridge, MA.}}
\bigskip

\address{
Department of Physics \\
University of Maryland \\
College Park, Maryland 20742 \\
{~}}

\date{U. of MD PP\#97-103 ~~~DOE/ER/40762-118~~~ April 1997}

\maketitle

\begin{abstract}
A class of purely-nonperturbative (PNP) composite 
operators is defined in Quantum Chromodynamics, 
which is perturbative scheme and scale independent 
and are useful to describe the internal structure of 
a strong interacting system. The operator
product expansion in terms of the new operators 
cleanly separates the perturbative and nonperturbative
physics without introducing any factorization
scale. A number of PNP observables of the nucleon
is briefly discussed including the PNP parton 
distributions. In particular, the fraction of the 
nucleon momentum carried by the purely-nonperturbative 
gluons is found to be around 16\%.

\end{abstract}
\pacs{xxxxxx}

\narrowtext

Composite operators composed of elementary 
fields are used frequently in quantum 
field theory \cite{collins}. They are particularly useful
in constructing effective field
theories or operator product expansions (OPE) when
multiple scales appear in a physical problem.
It appears that there are two classes of 
intrinsically different composite operators:
those that are associated with the conserved currents
arising from continuous symmetries of a quantized
lagrangian, such as the vector currents in 
Quantum Chromodynamics (QCD), and those that are not. The latter type
of operators, in their bare form, are typically 
divergent and require renormalization. The renormalized 
operators are known to have scheme and scale dependence. 
One argues that these operators are present 
only in intermediate stages of a calculation 
and the final physical results
are independent of any particular renormalization 
procedure.

Nonetheless, there are strong motivations to go beyond
the viewpoint that composite operators are 
just auxiliary quantities. In an OPE, one often 
views the composite operators 
describing the physics below the perturbative 
renormalization scale $\mu^2$. When $\mu^2$ is at 
the hadron mass scale, these 
operators act as a probe into the nonperturbative 
wave function of a hadron system. In particular, 
the nucleon matrix elements of composite operators 
are long regarded as useful observables exhibiting 
the structural information of the nucleon. For 
instance, the scalar charge of the nucleon 
$\langle P|\bar \psi \psi|P\rangle$, 
quark distributions, etc. are routinely calculated 
in nucleon models. 

However, scheme and scale dependence of 
composite operators makes the above interpretations
unconvincing. For instance, it is unknown 
how much physics in the matrix element of a 
renormalized operator $\hat O_R(\mu)$ is attributed to
perturbation theory and how much is truly nonperturbative, 
i.e., a reflection of the bound state properties. 
The conventional wisdom is that the perturbative 
contributions can be evolved away by going to small $\mu$.
In practice, no consensus has ever been reached as to 
precisely which. Strong sensitivity of matrix elements
in low scale $\mu$ renders any particular 
choice difficult to justify. Furthermore, a renormalized 
composite operator is arbitrary up to a multiplicative 
perturbation series 
$1+\sum_{i=1}^\infty c_n \alpha_s^n(\mu^2)$, which can be 
significantly different from unity when $\alpha_s(\mu)$
becomes large. A less severe, but still significant 
problem is that the experimentally-extracted 
matrix elements of composite operators, such as
parton distributions, cannot be seriously compared with
model calculations. In a nucleon model, the
physics associated with renormalization is often obscure; 
it is difficult to assign a renormalization scale and/or scheme to 
model matrix elements \cite{jaffe}.

In this Letter a new renormalization definition is offered for  
a class of logarithmically-divergent operators in QCD.
The newly-defined operators are perturbative
scheme and scale independent. Their matrix elements 
in a hadron state depend {\it only} on the strong
interaction scale $\Lambda_{\rm QCD}$ and therefore are 
totally nonperturbative. To distinguish them
from the conventionally-renormalized operators, we call
them ``purely-nonperturbative (PNP) operators". 
An interesting by-product is that the 
coefficient functions in a reformulated OPE with 
the PNP operators are also perturbative scheme 
independent.  

The Letter is organized as follows. First 
the new definition of the renormalized composite 
operators in QCD is presented. Then these 
operators are shown to be perturbative 
scheme and scale independent. 
Following that, their use in OPE is discussed and 
the corresponding coefficient functions are shown to be
also perturbative scheme independent. Subsequently, 
the new form of OPE and the traditional momentum factorization are 
compared. Finally, a number of comments 
are made about applications of the PNP operators. 
In particular, the notion of the PNP parton 
distributions is introduced.

Let us begin by recalling the standard definition
of composite operators in quantum field theory \cite{zimmermann}. 
For simplicity, we will not consider explicitly the operator 
mixing, for which the discussions below are unchanged
except all equations must be in matrix form.
Consider a bare composite operator $\hat O_B$,
and insert it into all the renormalized 
elementary Green's functions. Assume 
primitively-divergent Feynman diagrams have  
only overall logarithmic 
divergences, which have to be subtracted to define 
the renormalized operators,
\begin{equation}
      \hat O_R(\mu) = Z^{-1}(\mu) \hat O_B \ ,
\end{equation}
where $Z(\mu)$ is a perturbation series in $\alpha_s(\mu)$. 
{}From the viewpoint of factorization, 
$Z(\mu)$ factorizes out the contributions
to the bare operator from momentum scales  
larger than $\mu$. $\hat O_R(\mu)$
is far from unique because it not only depends on 
the scale $\mu$ but on choices of the finite part of
the renormalization constant $Z(\mu)$. 

The central point of this paper is that 
in QCD one can define a finite operator
$\hat O^{\rm pnp}$ from $\hat O_R(\mu)$ by 
making a further factorization:
\begin{equation}
     \hat O^{\rm pnp} = \Delta Z^{-1}(\mu) \hat O_R(\mu) \ ,
\end{equation}
where $\Delta Z(\mu)$ is defined as
\begin{equation}
    \Delta Z(\mu) = (\alpha_s(\mu))^{\gamma_0/2\beta_0}
               \exp\left(-\int^{\alpha_s(\mu)}_0
   \left( { \gamma(\alpha)\over \beta(\alpha)}+{\gamma_0\over \beta_0} \right) 
    {d\alpha \over 2\alpha} \right) \ , 
\end{equation}
where $\gamma(\alpha) = Z^{-1} \mu dZ/d\mu =\gamma_0\alpha/4\pi +...$ is the 
anomalous dimension of the operator and $\beta(\alpha) = -\beta_0
\alpha/4\pi+...$ $(\beta_0=11-2n_f/3)$ is the QCD 
beta-function. This additional,
perturbatively-calculable 
factorization takes away all the dependence 
on the scale $\mu$, introduced artificially in the perturbative
factorization in Eq. (1). $\Delta Z(\mu) Z(\mu)$ is 
$\mu$ independent as one can check by direct differentiation.
$\hat O^{\rm pnp}$ can have no other 
scale dependence except $\Lambda_{\rm QCD}$
and therefore is a purely-nonperturbative
operator. 
Notice that generally $\hat O^{\rm pnp}$ cannot be defined 
order-by-order in perturbation theory. In particular, its 
matrix element in any perturbative state takes the form,
\begin{equation}
     M^{\rm pnp} = \Delta Z^{-1}(\mu)\left[1-{\alpha_s(\mu)\over 8\pi}
         \left(\gamma_0\ln\left(
           {\mu^2\over m_q^2}\right)+c_0\right)+...\right]
\end{equation}
where $m_q$ is a soft quark mass and $c_0$ some constant. 

$\hat O^{\rm pnp}$ is also independent of perturbative 
factorization schemes. Assume that one has adopted a 
different factorization,
\begin{equation}
  \tilde {\hat O}_R =\tilde Z^{-1}(\mu^2)\hat O_B \ ,
\end{equation}
where $\tilde Z(\mu^2)$ differs from $Z(\mu^2)$ by 
a perturbation series $z(\alpha) = 1+ \sum_{i=1}^\infty z_i\alpha^i$.
The new anomalous dimension becomes
\begin{equation}
    \tilde \gamma(\alpha) = \gamma(\alpha) + 
 \mu{d\ln z(\alpha(\mu))\over d \mu}  \ .
\end{equation}
Substituting $\tilde \gamma(\alpha)$ in $\Delta \tilde Z$,
one has
\begin{equation} 
    \Delta \tilde Z^{-1}(\mu) {\tilde {\hat O}}_R(\mu) = \Delta Z^{-1}(\mu) 
\hat O_R(\mu)
      =\hat O^{\rm pnp} \ .
\end{equation}
Thus apart from an overall multiplicative numerical constant,
$\hat O^{\rm pnp}$ is uniquely defined. If one further imposes
the condition that for a conserved current, $O^{\rm pnp}$
must coincide with the bare operator, the multiplicative
constant is reduced to a function of $\gamma_0$, which 
takes the value 1 when $\gamma_0=0$, e.g., $(\lambda)^{\gamma_0/2\beta_0}$. 
This freedom is intrinsically different from
the ambiguity in perturbative factorization schemes, 
and is analogous to the arbitrariness in 
the choice of a unit for electric charge and the 
corresponding electromagnetic field. The ``natural'' convention, 
as we will see below, is the one we have taken for 
$\Delta Z$ in Eq. (3). One can obviously 
generalize the above discussion to the quark mass
renormalization. The result is the so-called renormalization
group invariant mass, which has been known 
in the literature for some time \cite{verm}.

The measurement of the matrix element of 
a PNP operator generally requires high momentum 
processes. Consider a physical observable $A(Q)$, where
$Q^2$ is a large physical scale, such as the mass of a
virtual photon. Assume $A(Q)$ has a conventional OPE 
with a leading term depending on $M(\mu)$, the 
matrix element of $\hat O_R(\mu)$,
\begin{equation}
     A(Q) = C(Q^2/\mu^2, \alpha_s(\mu)) M(\mu) + ... \ , 
\end{equation}
where $C$ is a coefficient function calculable as a 
perturbation series in $\alpha_s(\mu)$. [We ignore
the possibility that the $C$ perturbation series may be 
divergent due to infrared and ultraviolet renormalons 
or instanton-anti-instanton pairs \cite{mueller1}.] Clearly both $M(\mu)$ 
and $C$ are factorization scheme and scale dependent. 
However, because $A(Q)$ is a physical observable, 
the coefficient function $C$ obeys the following 
renormalization group equation,
\begin{equation}
        \mu {dC\over d\mu} -\gamma(\alpha_s(\mu))C =0 \ . 
\end{equation}
The solution of Eq. (9) may be written as,
\begin{eqnarray}
        C\left({Q^2/\mu^2}, \alpha_s(\mu)\right)
       &=& C(1, \alpha_s(Q)) \exp
        \left(-\int^{\alpha_s(Q)}_{\alpha_s(\mu)}
         {\gamma(\alpha)\over \beta(\alpha)}{d\alpha\over 2\alpha} \right) \nonumber \\
       &=& C(1, \alpha_s(Q))\left(\alpha_s(Q) 
       \right)^{\gamma_0/2\beta_0}
         {\exp \left(-\int^{\alpha_s(Q)}_0\left({\gamma(\alpha)\over 
        \beta(\alpha)}
         +{\gamma_0\over \beta_0}\right){d\alpha \over 2\alpha}
           \right) } \nonumber \\
         && \times
      \left(\alpha_s(\mu) \right)^{-\gamma_0/2\beta_0}
       \exp \left(\int^{\alpha_s(\mu)}_0\left({\gamma(\alpha)\over 
  \beta(\alpha)}
         +{\gamma_0\over \beta_0}\right){d\alpha\over 2\alpha} \right) \ . 
\end{eqnarray}
In the second line of the above equation, the $Q$
and $\mu$ dependence is completely factorized. 
The $\mu$-dependent part is ``naturally'' the 
$\Delta Z^{-1}$ defined in Eq. (3). Therefore we have,
\begin{equation}
       A(Q) = C^{\rm inv}(\alpha_s(Q)) M^{\rm pnp} + ...
\end{equation}
where $M^{\rm pnp}$ is the matrix element of 
$\hat O^{\rm pnp}$ in Eq. (4). Now $C^{\rm inv}$ is a
function of $\alpha_s(Q)$ only,
\begin{equation}
     C^{\rm inv}(\alpha_s(Q)) = C_0~(\alpha_s(Q))^{
               \gamma_0/2\beta_0}~
            \left(1+\sum_{n=1}^\infty c_n \alpha_s^n(Q)\right) \ , 
\end{equation}
and is perturbative scheme independent although
the coupling constant $\alpha_s(Q)$ is not.
This means that $C^{\rm inv}$ is as physical as 
the perturbation series for the total 
$e^+e^-\rightarrow {\rm hadron}$ cross section. 
Through this physical probe $C^{\rm inv}$,
the nonperturbative matrix element $M^{\rm pnp}$ can 
be extracted from the observable $A(Q)$ in the asymptotic limit,
\begin{equation}
          M^{\rm pnp} = \lim_{Q^2\rightarrow \infty}
           A(Q) /C^{\rm inv}(\alpha_s(Q)) \ .  
\end{equation}

Equation (10) represents a nonperturbative factorization: 
all probe-related ($Q^2$) physics 
is included in $C^{\rm inv}(\alpha_s(Q))$, which 
is generally nonperturbative in $\alpha_s(Q)$ ($\gamma_0\ne 0$), 
but is perturbatively calculable; all target-related
physics is in $M^{\rm pnp}$, which is purely-nonperturbative, 
ultraviolet safe, and unique. Emergence of the renormalization
group invariants from solving evolution equations of 
the coefficient functions has been noted 
before \cite{buras}; however, its significance, to the author's
knowledge, was not explored.

At this point it is useful to comment on the relation between
this new form of factorization and the traditional momentum 
factorization. If one computes $A(Q)$ in perturbation theory
in a quark-gluon state, mass singularities may arise
in the form of the following integral,
\begin{equation}
      \int^{Q^2}_{m^2_q} {dk^2_\perp\over k^2_\perp} \ ,  
\end{equation}
where $m_q$ is a quark mass. To isolate the contributions
from the small $k_\perp$ region, one usually splits the integral
by introducing a factorization scale $\mu$,
\begin{equation}
    \int^{Q^2}_{\mu^2}
            {dk^2_\perp\over k^2_\perp} +  
       \int^{\mu^2}_{m^2_q} {dk^2_\perp\over k^2_\perp} \ .  
\end{equation}
The first term contains hard momentum only and can be identified
as a part of the coefficient function $C$. The second term 
is soft and is identified as a part of the perturbative 
matrix element of a nonperturbative operator $\hat O_R(\mu)$. 
It seems that the asymptotic freedom is best respected in this 
way. According to the discussions earlier, however, the second term
still contains perturbatively calculable contributions. 
Indeed, a truly nonperturbative 
quantity, such as hadron mass, should not be sensitive to the
ultraviolet cutoff $\mu$. The present $\mu$ dependence in 
the second term can be further factorized away, 
leaving the purely-nonperturbative contribution, as
signalled by the mass singularity, intact. However,
this cannot be done order-by-order in perturbation
theory as $\Delta Z$ in Eq. (3) and the perturbative matrix elements
of the PNP operators in Eq. (4) are generally 
non-analytic in $\alpha_s(\mu)$. An order-by-order 
isolation of the mass singularity necessarily generates 
the perturbative $\mu$-dependence in the soft part. 
From a different perspective, one might view the momentum
factorization as the process of splitting the 
physical $C^{\rm inv}$ into a product of two 
perturbative factors $C(\mu)$ and 
$\Delta Z(\mu)$, and then combining the latter with the 
the matrix element in Eq. (4) to produce a perturbative 
expansion in $\alpha_s(\mu)$. To be sure, 
this might be a necessary intermediate step to 
calculate $C^{\rm inv}$. 

It is interesting to note that for a given conventional
factorization scheme and a given operator, one can always
find a $\mu_0$ such that $\Delta Z(\mu_0)=1$. At this $\mu_0$,
the above momentum factorization into perturbative and 
nonperturbative contributions seems effectively complete. 
Therefore, one could take $\mu_0$ as a matching
point between model calculations and data. 
Since $\mu_0$ is generally small ($\alpha_s(\mu_0)
\sim 1$), the actual perturbative evolution to 
such low scales is questionable.

The usefulness of the PNP operators 
lies in their applications to studying the internal
structure of hadrons. As we have said, the 
operators with given quantum numbers and 
dimension are essentially unique and completely 
nonperturbative; therefore, it is proper 
to interpret them as quasi observables.
On the practical side, the matrix elements of 
various operators calculated in nucleon models
are most naturally identified with those
of the PNP operators in QCD. [Of course, the issue 
of ``effective charges" still remains  
\cite{mueller2}.] The observables
measured in high energy processes, such as parton 
distributions, can now be totally decoupled from the 
high-energy aspects of the probes and attributed solely
to the internal structure of targets. Before 
closing, a few comments are made in this direction:

1) Parton distributions are normally defined in a 
scheme (usually $\rm \overline{MS}$) and scale.
One can now define purely-nonperturbative parton distributions
of hadrons with their moments equal to the matrix 
elements of the PNP twist-two operators. For many years,
Brodsky and his collaborators have studied the 
``intrinsic parton distributions" \cite{brodsky}. 
They assert that intrinsic distributions can be calculated 
in terms of the (light-cone) wave functions of the 
hadrons only and have a weak scale-dependence.
The intrinsic distributions may be identified with 
the PNP distributions. Compared with the ordinary 
parton distributions of a given scheme, 
the PNP distributions have a tamed 
small-$x$ behavior. Indeed, the $\Delta Z$ factor 
contributes to much of the small-$x$ rising in the 
ordinary parton distributions because $\gamma_0$ is a 
monotonically increasing function of the spin of 
the twist-two operators
and is singular or negative near spin-zero. 
The results for the PNP distributions, together with
other studies, will be presented in a separate paper \cite{ji2}.

2) According to standard perturbative calculations, gluons 
carry a fraction $16/(16+3n_f)$ of the nucleon momentum
and spin in the asymptotic limit \cite{gross}. Although this 
result is scheme independent also, it is completely perturbative 
as it is independent of hadron states.
The fraction of the momentum carried by the purely-nonperturbative 
gluons, $\pi^{\rm pnp}_g$, is related to the conventional 
fraction $\pi^{\rm\overline {MS}}_g(\mu)$ by ($n_f=3$),
\begin{equation}
        \pi_g^{\rm pnp} = 0.64 + \left[-0.64+0.12{\alpha_s^{\rm 
     \overline{MS}}(\mu)\over \pi}
        +\pi^{\rm \overline{MS}}_g(\mu)\left(1-0.57{\alpha_s^{\rm 
       \overline{MS}} (\mu)\over \pi}
       \right) \right]\left(\alpha_s^{\rm \overline{MS}}(\mu)\right)
        ^{-50/81}\  
\end{equation}
at the next-to-leading logarithmic order. 
Using $\pi_g^{\rm \overline{MS}}$(1.6 GeV) = 0.42 \cite{cteq4}
and $\alpha_s^{\rm \overline{MS}}$(1.6 GeV) = 0.3, one finds that the
gluons carry only 16\% of the nucleon momentum.  
This not only confirms that the PNP glue distribution 
is much reduced, but also lends important support for
valence quark pictures.

3) The fraction of the nucleon spin contributed by the spin
of the quarks $\Delta \Sigma(\mu)$ (singlet axial charge), defined 
in $\langle P|\sum_{i=u,d,s}\bar \psi_i\gamma^\mu\gamma_5 \psi_i
|P\rangle = 2\Delta\Sigma(\mu) S^\mu$, is scheme and scale dependent
because of the Adler-Bell-Jackiw anomaly. The purely 
non-perturbative fraction can be defined, 
\begin{equation}
   \Delta  \Sigma^{\rm pnp} = \exp\left(\int^{\alpha_s(\mu^2)}_0
      {\gamma(\alpha)\over \beta(\alpha)}{d\alpha\over 2\alpha}\right)
     \Delta \Sigma(\mu) \ ,
\end{equation}
where $\gamma(\alpha)$ starts at order $\alpha^2$. 
This definition of $\Sigma^{\rm pnp}$
has already been recognized before \cite{collins,larin}. 
The actual deviation of $\Delta \Sigma^{\rm pnp}$ from, say,
$\Delta \Sigma^{\rm \overline{MS}}(\mu^2=10~ {\rm GeV}^2)$ is 
small. The fraction of the nucleon spin carried
by the purely-nonperturbative gluons in the light-like gauge
$(A^+=0)$ is related to the conventional scheme and scale
dependent fraction by,
\begin{equation}
    \Delta G^{\rm pnp} = (1+c_1\alpha_s(\mu)+...)
  \alpha_s(\mu)(\Delta G(\mu) + 4\Delta\Sigma/beta_0)
\end{equation}
Using $\Delta G(1~{\rm GeV}) = 1.6\pm 0.9$ \cite{altarelli}, 
we get $\Delta G^{\rm pnp} = 0.72 \pm 0.4$ in the leading
order, a reduced but still significant number.  

4) In recent years, the so-called strange content 
of the nucleon has been studied extensively. While the
strange radius and anomalous magnetic moment of the nucleon,
defined from the strange vector current $\bar s\gamma^\mu s$,
are scale and scheme independent, other strange observables,
e.g., strange scalar, axial, and tensor charges, are not 
\cite{musolf}. Because of this, questions 
have often been raised whether the latter quantities 
can faithfully describe the strange content of the 
nucleon. One can now use the PNP operators to achieve this. 
For instance, the strange scalar charge defined in 
$\langle P|\bar ss|P\rangle = s(\mu^2) (2M)$
has a PNP version,
\begin{equation}
      s^{\rm pnp} = (\alpha_s(\mu^2))^{4\over 9}(1+
     0.90{\alpha_s(\mu^2) \over \pi}+...)s(\mu^2)\ . 
\end{equation}
which can be extracted from the $\pi-N$ $\sigma$ term.

\acknowledgements
I would like to thank V. Braun, S. Brodsky, T. Cohen, 
J. Collins, R. Jaffe, W. Melnitchouk, A. Mueller, E. Shuryak, 
X. Song, and A. Thomas for their useful discussions 
and critical comments. This work is supported in part 
by funds provided by the U.S.  Department of 
Energy (D.O.E.) under cooperative agreement
DOE-FG02-93ER-40762.


\begin{references}
\frenchspacing

\bibitem{collins}
For a systematic discussion on composite operators, see for
instance, J. C. Collins, {\it Renormalization}, Cambridge
University Press, Cambridge, 1984. 

\bibitem{jaffe}
R. L. Jaffe and G. G. Ross, Phys. Lett. {\bf 93B}, 313 (1980);
F. M. Steffens and A. W. Thomas, Prog. Theor. Phys. Suppl.
{\bf 120}, 145 (1995); D. Diakonov, V. Petrov, P. Pobylitsa, M. Polyakov, 
hep-ph/9606314, 1996.

\bibitem{zimmermann}
W. Zimmermann, Ann. Phys. (New York) {\bf 77}, 536; 570 (1973).

\bibitem{verm}
For a recent reference, see 
J. A. M. Vermaseren, S. A. Larin, T. van Ritbergen,
hep-ph/9703284, 1997.

\bibitem{mueller1}
A. H. Mueller, Phys. Lett. {\bf B308}, 355 (1993). 

\bibitem{buras}
A. J. Buras, Rev. Mod. Phys. {\bf 52}, 199 (1980); F. J. Yndurain,
{\it Quantum Chromodynamics}, Springer-Verlag, New York, 1983.

\bibitem{mueller2}
A. H. Mueller, {\it Proceedings of Twelfth International 
Conference on Particles and Nuclei}, ed. by J. L. Matthews et al., 
North-Holland, 1991.

\bibitem{brodsky}
S. J. Brodsky, P. Hoyer, C. Peteson, N. Sakai, Phys. 
Lett. {\bf 93B}, 451 (1980);
S. J. Brodsky, S. Peterson, N. Sakai, Phys. Rev. {\bf D23}, 2745 (1981);
S. J. Brodsky, I. A. Schmidt, Phys. Lett. {\bf B234}, 144 (1990); R. Vogt
and S. J. Brodsky, Phys. Lett. {\bf B349}, 569 (1995). 

\bibitem{ji2}
X. Ji and W. Melnitchouk, to be published.

\bibitem{gross}
D. Gross and F. Wilczek, Phys. Rev. D9, 980 (1974);
X. Ji, J. Tang, P. Hoodbhoy, Phys. Rev. Lett. {\bf 76}, 740 (1996).

\bibitem{cteq4} 
CTEQ Collaboration: H. L. Lai et al., Phys. Rev. {\bf D55}, 1288 (1996).

\bibitem{larin}
For recent references on this, see S. A. Larin, 
T. van Ritbergen, J. A. M. Vermaseren, hep-ph/9702435, 1997;
also S. D. Bass, R. J. Crewther, 
F. M. Steffens, A. W. Thomas, hep-ph/9701213, 1997.

\bibitem{altarelli}
G. Altarelli, R. D. Ball, S. Forte, and G. Ridolfi, 
hep-ph/9701289, 1997.

\bibitem{musolf}
A. Manohar and D. Kaplan, Nucl. Phys. {\bf B310}, 527 (1988);
R. L. Jaffe, Phys. Lett. {\bf B229}, 275 (1989); M. Musolf et. al., 
Phys. Rep. {\bf 239}, 1 (1994).

\nonfrenchspacing
\end{references}
\end{document}